\documentclass[a4paper,11pt]{article}
\pdfoutput=1
\usepackage[normalem]{ulem}
\usepackage{jcappub} 
\usepackage{amsmath}
\usepackage{amssymb}
\usepackage{color}
\usepackage{graphicx}
\usepackage{comment}
\usepackage{multirow}
\usepackage[dvipsnames]{xcolor}
\usepackage[utf8]{inputenc}
\usepackage{caption}

\include{journals}

\newcommand{\om}{\Omega_{\rm m}}

\newcommand{\ol}{\Omega_{\Lambda}}

\newcommand{\oko}{\Omega_{\rm K}}

\newcommand{\wm}{\omega_{\rm M}}
\newcommand{\wg}{\omega_{\gamma}}

\newcommand{\wk}{\omega_{\rm K}}
\newcommand{\wb}{\omega_{\rm B}}

\newcommand{\wrad}{\omega_{\rm Rad}}
\newcommand{\wde}{\omega_{\rm DE}}
\newcommand{\wede}{w_{\rm EDE}}

\newcommand{\mn}{{\mu\nu}}

\newcommand{\be}{\begin{equation}}
\newcommand{\ee}{\end{equation}}

\begin{document}
\title{\boldmath Does the Hubble constant tension call for new physics?} 
\author[a]{Edvard M{\"o}rtsell,}
\affiliation[a]{Oskar Klein Centre, Department of Physics, Stockholm University, SE 106 91 Stockholm, Sweden}
\emailAdd{edvard@fysik.su.se}

\author[a]{Suhail Dhawan}
\emailAdd{suhail.dhawan@fysik.su.se}

\abstract{
The $\Lambda$ Cold Dark Matter model ($\Lambda$CDM) represents the current standard model in cosmology. Within this, there is a tension between the value of the Hubble constant, $H_0$, inferred from local distance indicators and the angular scale of fluctuations in the Cosmic Microwave Background (CMB). We investigate whether the tension is significant enough to warrant new physics in the form of modifying or adding energy components to the standard cosmological model. 
We find that late time dark energy explanations are slightly disfavoured whereas a pre-CMB decoupling extra dark energy component has a marginally positive Bayesian evidence. A constant equation of state of the additional early energy density is constrained to 0.086$^{+0.04}_{-0.03}$. Although this value deviates significantly from 1/3, valid for dark radiation, the latter is not disfavoured based on the Bayesian evidence. 
If the tension persists, future estimates of $H_0$ at the 1$\%$ level will be able to decisively determine which of the proposed explanations is favoured. 
}
\maketitle
\section{Introduction}\label{sec-intro}
As this is written, it is more than a century since Albert Einstein \cite{Einstein:1917ce} and Willem de Sitter \cite{deSitter:1917zz} suggested that our Universe could be described in terms of the relativistic field equations proposed by Einstein two years earlier \cite{Einstein:1915ca}. 

In 1922, Alexander Friedmann presented solutions to the field equations, with matter and a time dependent scale factor \cite[English translation;][]{Friedmann:1924bb}. Friedmann's work was both rebutted and later unrebutted by Einstein, and the solutions were not generally acknowledged until a few years after Lema\^{i}tre's rediscovery in 1927 \cite{Lemaitre:1927zz}. Lema\^{i}tre also related the solutions to observed recession velocities of distant sources. The proper accreditation for the observational discovery of the universal expansion is not uncontroversial, see e.g. \cite{Way:2013ky} and references therein. Here, we only note that in the thirties, it was observationally proven that the Universe is expanding and that this is a natural outcome of general relativity. Acknowledgements for this discovery should include Vesto Slipher, Milton Humason, Henrietta Swan Leavitt, Knut Lundmark, George Lema\^{i}tre and Edwin Hubble. 

In 1970, Allan Sandage described cosmology as the search for two numbers \cite{Sandage1970}: The current rate of expansion ($H_0$) and the deceleration of the expansion ($q_0$). Since the latter is directly connected to the energy content of the Universe, it has attracted most attention in modern cosmology, especially with the realization that $q_0<0$, i.e., that the expansion rate is increasing with time \cite{Riess:1998cb,Perlmutter:1998np}. The Hubble constant has mostly been considered important for being inversely proportional to distances and the age of the Universe. With the advent of precise CMB observations \cite{Planck2015}, it has been realized that $H_0$ is important also for measuring the energy content of the Universe.

$H_0$ can be both measured locally and derived from the angle subtended by the sound horizon as observed in CMB temperature fluctuations. Nature thus provides two absolute distance scales at opposite ends of the visible expansion history of the Universe. 

Improved measurements (3--5$\%$ precision) of $H_0$ at low redshifts \cite[$z \lesssim 0.5$; e.g.][]{Riess2009,Riess2011,Freedman2012,Suyu2013,Bonvin2017}, along with recent progress in CMB observations
\citep{Bennett2013,Hinshaw2013,Planck2015}, hinted at a mild tension ($2-2.5\,\sigma$) between the local and CMB measurements. 
The most precise current estimates of distances to local supernovae of type Ia (SN~Ia) come from observations of Cepheid variables in host galaxies of 19 SNe~Ia from the SH0ES program \cite{Riess2016}. Combined with a large SN~Ia sample up to $z\approx 0.15$, this results in a $H_0$ value with 2.4\,\% uncertainties, in 3.4$\,\sigma$ tension with the CMB value from \emph{Planck} \cite{Planck2016}.
Recent reanalyses of the data in \cite{Riess2016} confirm the insensitivity of $H_0$ to Cepheids systematics \cite{Follin2017}, SNe~Ia \cite{Dhawan2017c} and the model for fitting the data \cite{Feeney2017}, substantiating the possibility that the  tension is due to non-standard physics. 

In \cite{Planck2015}, it is noted that CMB data give only an indirect and highly model-dependent estimate of $H_0$. From figure~3 in \cite{Ade:2015rim}, it is clear that CMB data alone prefers phantom dark energy and high $H_0$ (the likelihood is still increasing at $H_0\sim 90$ km/s/Mpc). Lower $H_0$ is only favoured when including low redshift information in the form of Baryon Acoustic Oscillation (BAO) and SN~Ia data. 

\cite{Bernal2016} consider changes in the early time physics and reconstruct the late time expansion history using BAO and SN~Ia data. They find that dark radiation with an additional effective number of species around 0.4 could relieve the Hubble tension, but note that preliminary Planck CMB polarization data disfavours this solution. At low redshifts ($z\le 0.6$), the recovered expansion history deviates less than 5\,\% from the $\Lambda$CDM model.

\cite{Verde:2016wmz} explore how much additional energy can be placed in components beyond those in the flat $\Lambda$CDM model at early epochs. At 95\,\% confidence level (CL), additional tracking dark energy is constrained as $\Omega_{\rm add}<0.006$ and dark radiation, parameterised with the effective number of neutrino species, as $2.3 < N_{\rm eff} < 3.2$.

In this study, we investigate what the apparent tension implies in terms of new physics, both before and after CMB decoupling when relaxing the assumption of the $\Lambda$CDM cosmological model.  
Specifically, assuming that neither the local or CMB inferred values of $H_0$ are contaminated by major systematic effects, we employ a Bayesian analysis to study whether additional energy contributions are favoured by data. 

We use two different approaches in this study. First, we allow for the dark energy contribution to deviate from a constant density, i.e., a cosmological constant $\Lambda$. The equation of state (EoS) of dark energy is defined as the ratio $w=p/\rho$, with $p$ being the pressure and $\rho$ the energy density. For a cosmological constant, $w_\Lambda=-1$.
Previous work have investigated phenomenological dark energy EoS, e.g., 
\be\label{eq:wz}
w(z)=w_0+w_a(1-a)
\ee 
or expressing the Hubble parameter $H(z)$ in piece-wise natural cubic splines \cite{Bernal2016}. Here, we extend the phenomenological model of dark energy to maximize its possibility to resolve the Hubble tension, and also study an example of a fundamental theoretical model not well described by equation~(\ref{eq:wz}). Ghost-free bimetric gravity allows for non-monotonic EoS as well as negative effective dark energy densities. We also investigate additions to the $\Lambda$CDM model, e.g., in the form of early dark energy with an arbitrary constant EoS\footnote{In this paper, we use the term dark energy also to describe energy sources with $w\ge 0$.}.

We will make extensive use of the dimensionless Hubble constant expressed in units of 100 km/s/Mpc
\be
h\equiv\frac{H_0}{\rm 100\, km/s/Mpc}.
\ee
We define dimensionless densities by 
\be
\Omega_i=\frac{8\pi G\rho_i}{3H_0^2}\propto \frac{\rho_i}{h^2},
\ee
and frequently refer to physical densities in dimensionless form as $\omega_i\equiv \Omega_i h^2$. The velocity of light is set to unity, $c=1$.





\section{Methodology and Data}\label{sec-data}
A comparison between local and CMB measurement of $h$ can give important insights into dark energy properties \cite{Freedman2012,Riess2011,Riess2016}. The angular scales of CMB temperature anisotropies involves two length scales: The physical size of the sound horizon at the decoupling of CMB photons and the angular diameter distance to the decoupling redshift, $z_*$. The inferred value of the Hubble constant depends on the assumptions made when calculating these quantities. 

The sound horizon depends on the time of decoupling, as well as the sound speed and expansion rate before decoupling. For example, adding additional massless neutrino species increases the Hubble expansion rate, which decreases the sound horizon size and increases the inferred Hubble constant (in order to keep the observed angular sound horizon size constant). 

The angular diameter distance depends on, e.g., the EoS of dark energy. Decreasing $w$ increases the angular diameter distance which can be compensated by increasing the value of $h$. Since the observed CMB anisotropies are only sensitive to the integrated distance to the decoupling redshift, they are not sensitive to a possible temporal evolution of $w$. The observed angular size of the sound horizon points to a high value of $h$ and phantom dark energy with $w<-1$ \cite{Ade:2015rim}. A low $h$ is only preferred if dark energy is assumed to be in the form of a cosmological constant ($w_\Lambda=-1$), or when additional probes of the dark energy EoS are included, such as SNe~Ia \cite{Betoule2014} and BAO \cite{Alam2016} observations.
In this section, we describe the data employed in this paper and how they constrain the Hubble constant and the evolution of the expansion rate. 

To summarize, assuming $\Lambda$CDM, the local distance ladder constrains the Hubble constant using data out to $z\le 0.15$ to $h\sim 0.7324\pm 0.017$. Gravitational lensing probes the expansion out to $z\le 1.7$ giving $h\sim 0.72\pm 0.03$. CMB probes distances out to to $z\sim 1090$ giving $h\sim 0.6781\pm 0.0092$. Since the methods probe different distance ranges, the measurements can be made consistent if, e.g., the local expansion rates varies isotropically on large scales as in Lema\^{i}tre-Tolman-Bondi (LTB) models \cite{1934PNAS...20..169T,1947MNRAS.107..410B,1997GReGr..29..641L,GarciaBellido:2008nz,2010JCAP...05..006B,Sundell:2015cza}. We expect smaller scale inhomogeneities to have little impact on the derived parameters \cite{Dhawan:2017kft}. In this paper, we employ the cosmological principle and assume that the Universe is homogeneous and isotropic on large scales.

\subsection{Local measurements}
Local measurements of the Hubble constant refer to observations of the recession velocity of objects as a function of their distance, as in the original discovery of the universal expansion. Distance probes should cover a range large enough for objects at the far end of the Hubble diagram to have negligible peculiar velocities as compared to the Hubble flow. On the other hand, distances should be small enough for the impact of the time variation in the expansion rate to be insensitive to the assumed cosmology, restricting data to $z\lesssim 0.15$. The method is thus insensitive to the evolution of the Universe at $z\gtrsim 0.15$. However, it does depend on each step of the distance ladder not having large, unknown systematics.

In \cite{Riess2016}, four geometric distance calibrations of Cepheids are considered, giving slightly different estimates of $h$:
\begin{itemize}
\item The water megamaser in NGC$\,$4258: $h=0.7225 \pm 0.0251$,
\item 8 detached eclipsing binaries (DEBs) in the large Magellanic cloud: $h=0.7204 \pm 0.0267$,
\item 15 Milky way Cepheids with measured parallaxes: $h=0.7618 \pm 0.0237$,
\item 2 DEBs in M31: $h=0.7450 \pm 0.0327$,
\end{itemize}
yielding together a best estimate of $h=0.7324 \pm 0.0174$.

\cite{Dhawan2017c} derive $h=0.728 \pm 0.016\pm 0.027$ (statistical and systematic uncertainties), using 9 nearby calibrator and 27 SNe~Ia in the Hubble flow observed in the near-infrared (NIR), where intrinsic variations and extinction by dust are reduced relative to optical observations. This result suggest that SN~Ia systematics expected to vary between optical and NIR wavelengths, like dust extinction, have little impact on the derived $h$. 
For possible issues with biases from SNe~Ia having locally star-forming environments (in which Cepheids originate predominately) being dimmer than SNe~Ia having locally passive environments, see \cite{Rigault:2014kaa}.

\subsection{Gravitational lensing measurements}
The possibility to measure the Hubble constant using time delays between gravitationally lensed multiple images was first proposed by Refsdal in \cite{1964MNRAS.128..307R}. Refsdal was primarily envisioning SN~Ia sources
\cite[for the recent first such detected event, see][]{Goobar:2016uuf}, but also mentioned the possibility of the then recently discovered star-like objects with intense emission both at radio and optical wavelengths, today classified as quasars. 
Part of the COSMOGRAIL project, HOLiCOW \cite{Bonvin2017} uses three multiply-imaged quasar systems with measured gravitational time delays to estimate $h$. In flat $\Lambda$CDM, $h=0.719^{+0.024}_{-0.030}$, but the method is rather insensitive to the assumed cosmological model. 
Given properly measured time delays and positions, uncertainties are almost exclusively determined by uncertainties in the mass distribution of the lensing galaxy and possibly other masses close to the line of sight. If matter from lens and line-of-sight environments are ignored, the resulting $h$ would be overestimated by $11^{+3}_{-2}\,\%$ on average. If only line-of-sight groups are ignored, $h$ would be overestimated by $7^{+3}_{-2}\,\%$ \cite{Wilson:2017apg}.

\subsection{CMB measurements}
Temperature fluctuations in the CMB are determined by, e.g., the matter density $\wm$, the baryon density $\wb$, the angular distance to recombination $d_A(z_*)$ and the overall spectral tilt of primordial fluctuations, $n$ \cite{Hu:2001bc}. From these, other secondary parameters, including the Hubble constant can be derived. 

The observed angular scale of CMB fluctuations are primarily determined by two physical scales, the sound horizon at decoupling, $r_s(z_*)$ and the angular distance $d_A(z_*)$, where $z_*\approx 1090$. The former depends on sectors dominating the energy budget at $z>z_*$ (dark matter, baryons and radiation), and the latter on sectors dominating at $z<z_*$ (dark matter, baryons, dark energy and spatial curvature).

Following \cite{Efstathiou:1998xx}, the sound speed is given by
\be\label{eq:cs}
c_s= \frac{c}{\sqrt{3\left(1+\frac{3\wb}{4\wg}a\right)}},
\ee
and the sound horizon by
\be
r_s(z_*)=\int_0^{t_*}\frac{c_s dt}{a(t)}=\int_{z_*}^{\infty}\frac{c_s dz}{H(z)}=\int_{0}^{a_*}\frac{c_s da}{a^2H(a)}=\frac{1}{\sqrt{3}}\int_{0}^{a_*}\frac{da}{a^2H(a)\sqrt{1+\frac{3\wb}{4\wg}a}},
\ee
where $\wg$ is the energy density of radiation coupling electromagnetically to matter.
The comoving angular distance, $D_A\equiv (1+z)d_A$, is
\be
D_A=\frac{1}{\sqrt{-\oko}}\sin\left(\sqrt{-\oko}\int_0^{z}\frac{dz}{H(z)}\right).
\ee

Much of the information contained in the full CMB power spectrum can be compressed into two, so called shift parameters. The first, $\mathcal{R}$, is defined by
\be
\mathcal{R}=\sqrt{\wm}D_A(z_*),
\ee
the angular distance to $z_*$ divided by the comoving Hubble horizon at decoupling $R_H(z_*)=(1+z_*)/H(z_*)\sim \sqrt{1+z_*}\sqrt{\om}/H_0$, scaled with a factor of $\sqrt{1+z_*}$.
The second shift parameter, $l_A$, basically the angular distance to $z_*$ divided by the sound horizon at decoupling, is given by
\be
l_A=\pi\frac{D_A(z_*)}{r_s(z_*)}.
\ee
CMB anisotropies will look the same (at small scales) if (initial fluctuations being equal), both $\mathcal{R}$ and $[\wb,\wm]$ are unchanged\footnote{On scales $l>30$ it is broken by the late Sachs-Wolfe effect.}. This is the so called {\it geometrical degeneracy}. Given that CMB only constrains $\wb$ and $\wm$ and, say $\wk$ from $\mathcal{R}$, we can write 
\be
h^2=\frac{\wb+\wm+\wk}{1-\ol},
\ee
i.e., to constrain the Hubble constant, we need independent constraints on the cosmological constant. Allowing for more general dark energy properties causes further degeneracies.

In this work, we use the CMB compressed likelihood with three parameters $\mathcal{R}$, $l_A$ and $\Omega_b h^2$.
The value for ($R$,$l_A$, $\Omega_b h^2$) = (1.7382, 301.63, 0.02262) with errors (0.0088,0.15,0.00029) and covariance is
\begin{equation}
 D_\text{cmb} = \left(
\begin{array}{ccc}
1.0 &  0.64 &  -0.75 \\
0.64 & 1.0 & -0.55 \\
 -0.75 & -0.55 & 1.0 \\
\end{array}
\right)\,,
\label{eq:cmb_covar_pl15}
\end{equation}
such that the elements of the covariance matrix $C_{ij} = \sigma_i \sigma_j D_{ij}$. 

The redshift of decoupling is given by \cite{1996ApJ...471..542H}
\be
z_* = 1048 \cdot [1 + 0.00124\wb^{-0.738}][1+g_1 \wm^{g_2}],
\ee
where
\begin{align}
g_1 &= \frac{0.0783\wb^{-0.238}}{1+39.5\wb^{0.763}},\\
g_2 & = \frac{0.560}{1+21.1\wb^{1.81}}.
\end{align}
In the $\Lambda$CDM model, CMB data from \emph{Planck} constrains the Hubble constant to $h\sim 0.6781\pm 0.0092$. Including SN Ia and BAO data reduces the uncertainty further, giving $h\sim 0.6790\pm 0.0055$ \cite{Planck2015}. 

CMB constraints are quite insensitive to any astrophysical systematics, but highly dependent on the assumed energy content of the Universe, both before and after decoupling. 

\subsection{Type Ia supernovae}
In this section, SNe~Ia are discussed not in the sense of being used for local Hubble constant measurements, but as a probe constraining the universal expansion rate, when measuring $h$ using CMB observations.  
We employ the distance-redshift relation from the ``Joint Lightcurve Analysis" \cite[JLA,][]{Betoule2014} using their binned catalog of SN~Ia distances, following the procedure in \cite{2015PhRvD..92l3516A}. 
The distance modulus is calculated from the SN~Ia peak apparent magnitude ($m_B$), light curve width ($x_1$) and colour ($c$)
\be
\mu_{\rm obs} = m_B - (M_B - \alpha x_1 + \beta c),
\ee
where $M_B$ is the absolute magnitude of the SN~Ia. Following \cite{Betoule2014}, we apply a step correction for the host galaxy stellar mass. 
Now,
\be
\chi_{\rm SN }^2 = \Delta^T C_{\rm SN }^{-1} \Delta, 
\ee
where $\Delta = \mu - \mu_{\rm obs}$ and $C_{\rm SN }$ is the covariance matrix for the binned distances described in \cite{Betoule2014}.

\subsection{Baryon acoustic oscillations}
We have discussed how the angular size of the sound horizon at decoupling as imprinted in the CMB can be used to constrain cosmological parameters, including the Hubble constant. The same (redshifted) physical scale is also imprinted in the large scale distribution of matter and can be observed as percent level fluctuations, BAOs, in the matter density as probed by galaxies at $z\lesssim 1$ and the Ly-$\alpha$ forest at $z\sim 2$. In combination with CMB observations,
BAO data can be used to either infer cosmological parameters independently of $r_s$ and $h$ \cite{2009ApJ...703.1374S,Dhawan2017b}, or to break the degeneracy of $h$ as inferred from CMB observations with, e.g., dark energy properties. In this paper, we employ the latter approach.

We use a spherical average constraining a combination of the angular scale and redshift separation
\be
d_z = \frac{r_s(z_{\rm d})}{D_V(z)}, 
\ee
with 
\be
D_V(z) = \left[D_A(z)^2 \frac{z}{H(z)} \right] ^{1/3}. 
\ee
The drag epoch $z_{\rm d}$ is the time\footnote{The redshift is $z_{\rm d}=1059$ compared to $z_*=1090$, i.e. the baryons are decoupled $\sim$19\,000 years later than the photons.} when baryons are released from the
Compton drag of the photons. It is can be calculated using \cite{1998ApJ...496..605E}
\begin{align}
z_{\rm d} & = 1291 \frac{\wm^{0.251}}{1 + 0.659 \wm^{0.828} }
[1 + b_1 \wb^{b_2}],\\
b_1 & =0.313 \wm^{-0.419} [1 + 0.607 
\wm^{0.674} ], \\
b_2 & = 0.238 \wm^{0.223}.
\end{align}
We follow \cite{Planck2015} and use four measurements from 6dFGS at $z_{\rm eff} = 0.106$, the recent SDSS main galaxy (MGS) at $z_{\rm eff}$ = 0.15 of \cite{2015MNRAS.449..835R} and $z_{\rm eff}$ = 0.32 and 0.57 for the Baryon Oscillation Spectroscopic Survey (BOSS) \cite{2014MNRAS.441...24A}.
We consider a BAO prior of the form 
\begin{equation}
\chi^2_{\rm BAO} = (d_z - d_z^{\rm BAO})^T C_{\rm BAO}^{-1} (d_z - d_z^{\rm BAO}), 
\label{eq:chi_bao}
\end{equation}
with 
\begin{align}
d_z^{\rm BAO}& = [0.336, 0.2239,0.1181, 0.07206],\\
C_{\rm BAO}^{-1}& = {\rm diag}(4444.44,14071.64,183411.36,2005139.41). 
\end{align}

\subsection{Big bang nucleosynthesis}\label{sec:BBN}
The abundances of light elements from Big Bang Nucleosynthesis (BBN) are sensitive to the expansion velocity at  $0.2\cdot 10^9 < z < 3.7\cdot 10^9$.\footnote{Corresponding to 60 keV $< T <$ 1 MeV and 1 s $< t < 3$ min.} When investigating additional early dark energy models, e.g., dark radiation, we need to make sure that they are compatible with observational BBN constraints. 

If the expansion velocity is increased, equilibrium, i.e., the interaction rate $\Gamma\propto a^{-3}\propto t^{-3/2}$ being larger than the inverse age of the Universe $H\propto t^{-1}$, ends sooner and neutrinos decouple at higher temperature. The ratio of neutron, $n_n$, and proton densities, $n_p$, is
\be
\frac{n_n}{n_p}=\exp{\left(-\frac{1.29\,{\rm eV}}{T}\right)}.
\ee
Higher $T$ implies higher neutron fraction giving more helium and deuterium. Constraints from observed deuterium and helium fractions are complementary in constraining the dark radiation density and the baryon-to-photon ratio $\eta$, see e.g. \cite{Sasankan:2017eqr}. In this reference, it is argued that BBN constrains the amount of dark radiation at 10 Mev to be within -12.1\,\% and +6.2\,\% of the total background energy density, see also \cite{Verde:2016wmz}. These limits accommodate the energy densities needed to explain the Hubble tension, at least for any early dark energy EoS, $w\le 1/3$. 


\section{Late Dark Energy}
We begin by investigating whether it is possible to relieve the Hubble tension by allowing for the universal expansion rate to deviate from that of the $\Lambda$CDM model after CMB decoupling, i.e., at $z<z_*\approx 1090$. In doing this, we derive the preferred value of the Hubble constant for each model using CMB, BAO and SN~Ia data, as well as the locally derived value of $h$. If this global value of the Hubble constant is close to the locally derived value, the model can be considered efficient in relieving the Hubble tension. The efficiency can be quantified by calculating the the Bayesian evidence, see Section~\ref{sec-bayesian}.  

As useful points of reference, the combined CMB/BAO/SN~Ia limit for the $\Lambda$CDM model with the local Hubble prior corresponds to $h=0.697 \pm 0.0057$. For a constant dark energy EoS, we obtain $w=-0.98\pm 0.04$ and $h=0.695 \pm 0.008$.

\subsection{Phantom Dark Energy}
Phantom energy refers to an energy component with $w<-1$, for which the density grows with time in an expanding universe [and thus also $H(z)$ when the phantom component dominates the energy budget]. Such an energy component is of interest for the Hubble tension problem, since CMB data alone prefers high $h$ and phantom dark energy \cite{Ade:2015rim}.

Phantom energy would make the universe end in a Big Rip \cite{Caldwell:1999ew}, except possibly if $w\to -1$ in the infinite future \cite{Frampton:2011sp,Mortsell:2017fog}. It violates the dominant energy condition $\rho\ge |p|$ \cite{hawking1973large}, i.e., comoving observers may measure a negative energy density and matter may move faster than light. Derived from a canonical scalar field, the Hamiltonian has negative kinetic energy and is thus susceptible to vacuum instabilities \cite{Barenboim:2017sjk}. 
Masses of black holes decrease in a phantom energy background, tending to zero in a phantom energy universe approaching the Big Rip \cite{Babichev:2004yx} \citep[however, see also][]{Gao:2008jv}\footnote{For the relation to Hawking radiation, see \cite{2010PhLB..683....7E,Guariento:2007bs}.}. There are thus many reasons to be cautious about phantom energy models. However, as will become evident in the next sections, models exist for which an {\em effective} energy density displays phantom behaviour without any of the aforementioned problems.

\subsection{Bimetric gravity}\label{sec:bimetric}
Ghost-free bimetric gravity is described by the Lagrangian
\begin{align}
\label{eq:HRaction}
\mathcal{L}=&-\frac{M_{g}^{2}}{2}\sqrt{-\det g}R_{g}-\frac{M_{f}^{2}}{2}\sqrt{-\det f}R_{f}\nonumber \\
&+m^{4}\sqrt{-\det g}\sum_{n=0}^{4}\beta_{n}e_{n}\left(\sqrt{g^{-1}f}\right)+\sqrt{-\det g}\mathcal{L}_\mathrm{m},
\end{align}
where $\mathcal{L}_m$ is the matter Lagrangian and $e_n$ are elementary symmetric polynomials \cite{Hassan:2011vm}. Varying $\mathcal{L}$ with respect to the metrics $g_\mn$ and $f_\mn$, yields the equations of motion
\begin{align}
G^g_{\mu\nu}+m^{2}\sum_{n=0}^{3}\left(-1\right)^{n}\beta_{n}g_{\mu\lambda}Y_{\left(n\right)\nu}^{\lambda}\left(\sqrt{g^{-1}f}\right)&=\frac{1}{M_g^2}T_{\mu\nu},\\
G^f_{\mu\nu}+m^{2}\sum_{n=0}^{3}\left(-1\right)^{n}\beta_{4-n}f_{\mu\lambda}Y_{\left(n\right)\nu}^{\lambda}\left(\sqrt{f^{-1}g}\right)&=0.
\end{align}
Here, $M_f=M_g$ through a rescaling of $f_\mn$ and the $\beta_n$ \cite{Hassan:2011vm, Akrami:2015qga}. 

The specific metric interaction potential in the bimetric Lagrangian (\ref{eq:HRaction}), is designed to guarantee the absence of the so called Boulware-Deser ghost, corresponding to a propagating scalar degree of freedom. However, it is still possible to have other ghosts in the theory. In \cite{Higuchi:1986py}, it was shown that the helicity-$0$ mode of a massive spin-$2$ field in a de Sitter background will be ghost like if $0<m^2<2H^2$. The corresponding bound to avoid the Higuchi ghost in general isotropic and homogeneous backgrounds was derived in \cite{Fasiello:2013woa}. As shown in \cite{Konnig:2015lfa}, solutions fulfilling the Higuchi bound, exhibits a phantom behaviour of the effective dark energy density. 

From now, we define
\begin{equation}
B_i\equiv\frac{m^2 \beta_i}{H_0^2},
\end{equation}
and investigate combinations with two non-zero $B$-parameters.

\subsubsection{Linear model}
\begin{figure}
\centering
\includegraphics[width=11cm]{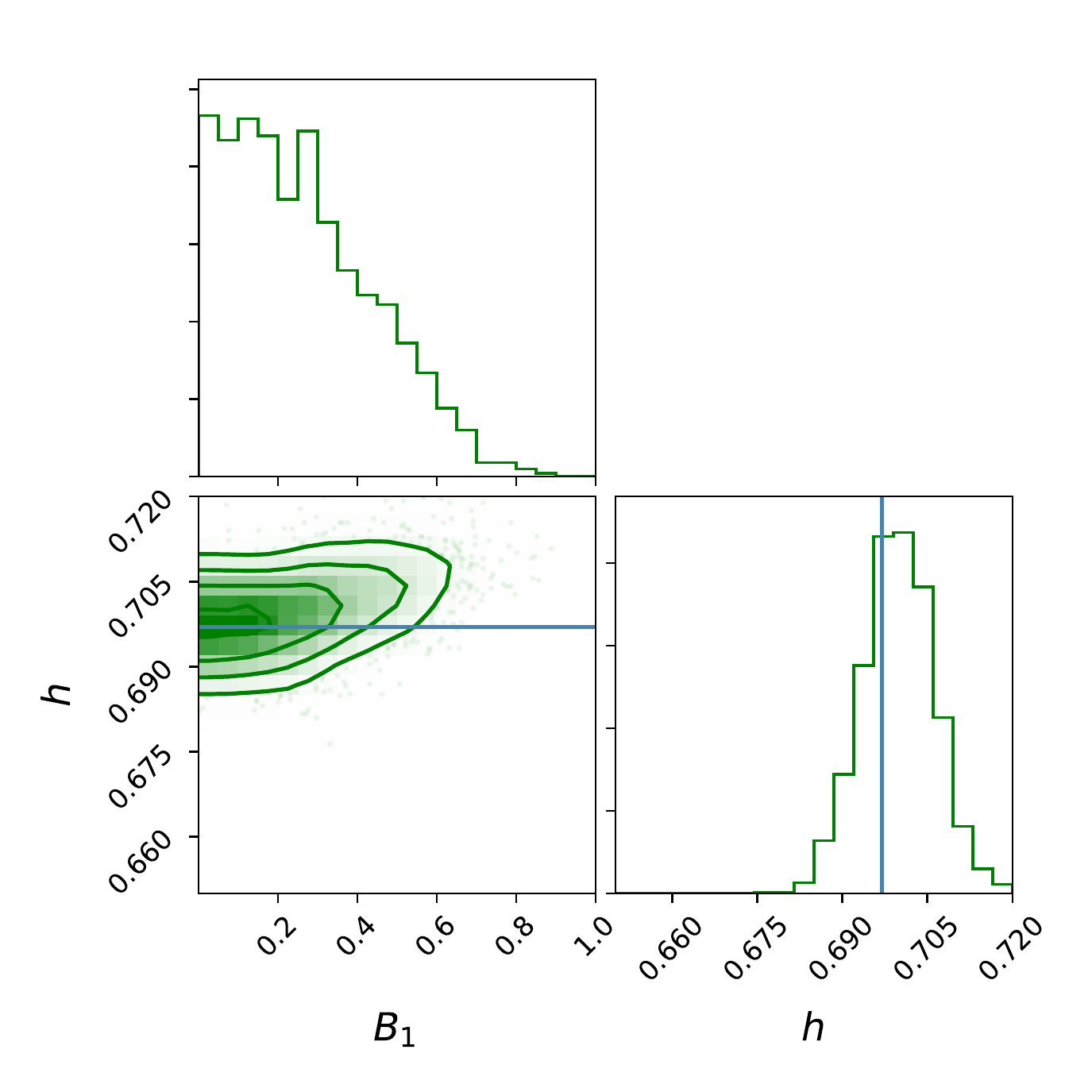}
\caption{\label{fig:triang_cmbbaosn_bimetricb0b1_h0} CMB/BAO/SN~Ia constraints on the linear bimetric model with $B_0$ and $B_1$, including the local Hubble constant prior. Observational data pushes $B_1\to 0$, corresponding to the $\Lambda$CDM limit of the model, giving $h=0.695\pm 0.006$, very similar to the $\Lambda$CDM value for the same combination of data, indicated by the vertical blue line.} 
\end{figure}

We first keep only terms in the interaction potential up to linear order, i.e., $B_0$ and $B_1$. The zeroth order term, $B_0$, dominates over the first order term at high redshifts. 
Asymptotically, the effective dark energy EoS, $w_r$, is time dependent and goes to $-1$ in the past and future infinity. 
Except for having an effective phantom dark energy EoS, this model is interesting since for $B_0<0$, we can have an effective negative dark energy density at redshifts beyond SN~Ia and BAO observations, which in principle can lower the Hubble expansion rate enough to relieve the Hubble tension. Figure~\ref{fig:triang_cmbbaosn_bimetricb0b1_h0}, however, shows that data pushes $B_1\to 0$, corresponding to the $\Lambda$CDM limit of the model and $h=0.695\pm 0.006$, very similar to the $\Lambda$CDM value.  

\subsubsection{Quadratic model}
We next include also the quadratic term, $B_2$, but set $B_0=0$. To have a real valued theory, $B_2<3/2$. This model is known to give a good fit when using the CMB/BAO ratio (being insensitive to the locally measured Hubble constant) and SN~Ia data since it has a well-defined $\Lambda$CDM limit when $B_2\to -\infty$ \cite{Dhawan2017b}. 
For $B_1\to 0$, we obtain a de Sitter model with $B_2=\Lambda=1$. 

\begin{figure}
\centering
\includegraphics[width=11cm]{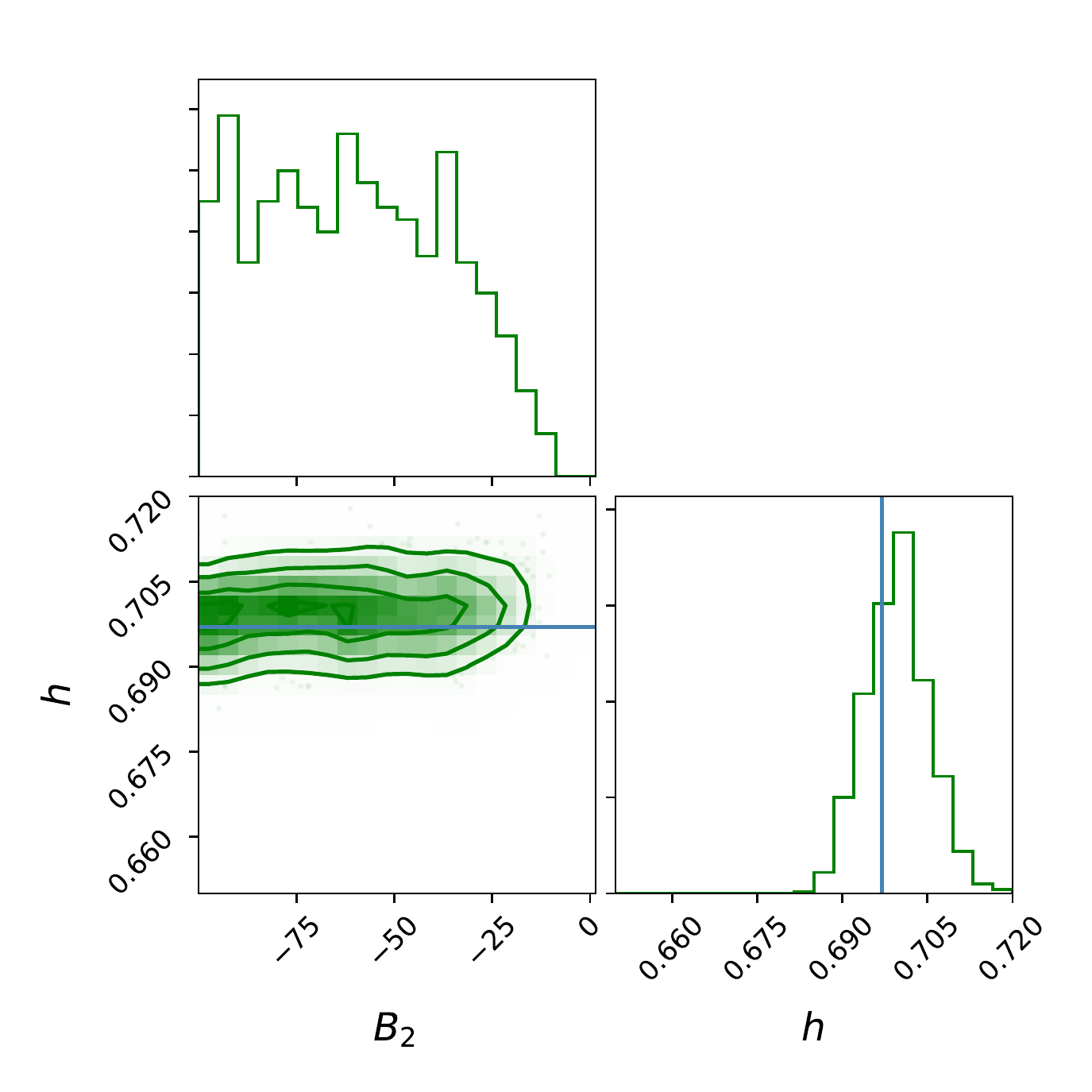}
\caption{\label{fig:triang_cmbbaosn_bimetric_h0_B1B22} CMB/BAO/SN~Ia constraints on quadratic bimetric model with $B_1$ and $B_2$, including the local Hubble constant prior. Observational data pushes $B_2\to -\infty$, corresponding to the $\Lambda$CDM limit of the model, giving $h=0.695\pm 0.006$, very similar to the $\Lambda$CDM value, indicated by the vertical blue line, for the same combination of data.}
\end{figure}

Since the effective dark energy EoS approaches $-2$ in the early matter dominated epoch (and $-7/3$ when radiation dominates), it is of interest to investigate possible implications for the Hubble tension. 
Only fitting for CMB data, higher values of the Hubble constant and phantom dark energy are in fact preferred. 
However, as evident from figure~\ref{fig:triang_cmbbaosn_bimetric_h0_B1B22}, including low redshift BAO and SN~Ia data pushes $B_2\to -\infty$, corresponding to the $\Lambda$CDM limit of the model. The Hubble constant is $h=0.695\pm 0.006$, identical to the allowed range in the bimetric linear model and very close to the $\Lambda$CDM result. 

\subsection{Phenomenological Dark Energy}
Phantom energy can help solve the Hubble tension by becoming negligible faster than $\Lambda$ when going back in time, thus decreasing the expansion rate and increasing the distance to $z_*$. However, phantom energy is in tension with SN~Ia and BAO data at $z\lesssim 1$ that pushes the dark energy EoS close to $w=-1$. It is natural to ask whether it is possible to relieve the Hubble tension through a modification of the expansion rate at redshifts above the SN~Ia and BAO data, but below CMB decoupling, i.e., at $1\lesssim z \lesssim 1090$. Since the $\Lambda$CDM model is matter dominated in this redshift interval, it is challenging to decrease $H(z)$ by modifying dark energy characteristics, since any additional positive dark energy density will slow the expansion down. Here, we investigate two approaches. 

\subsubsection{(Dis)appearing dark energy}
\begin{figure}
\centering
\includegraphics[width=15cm]{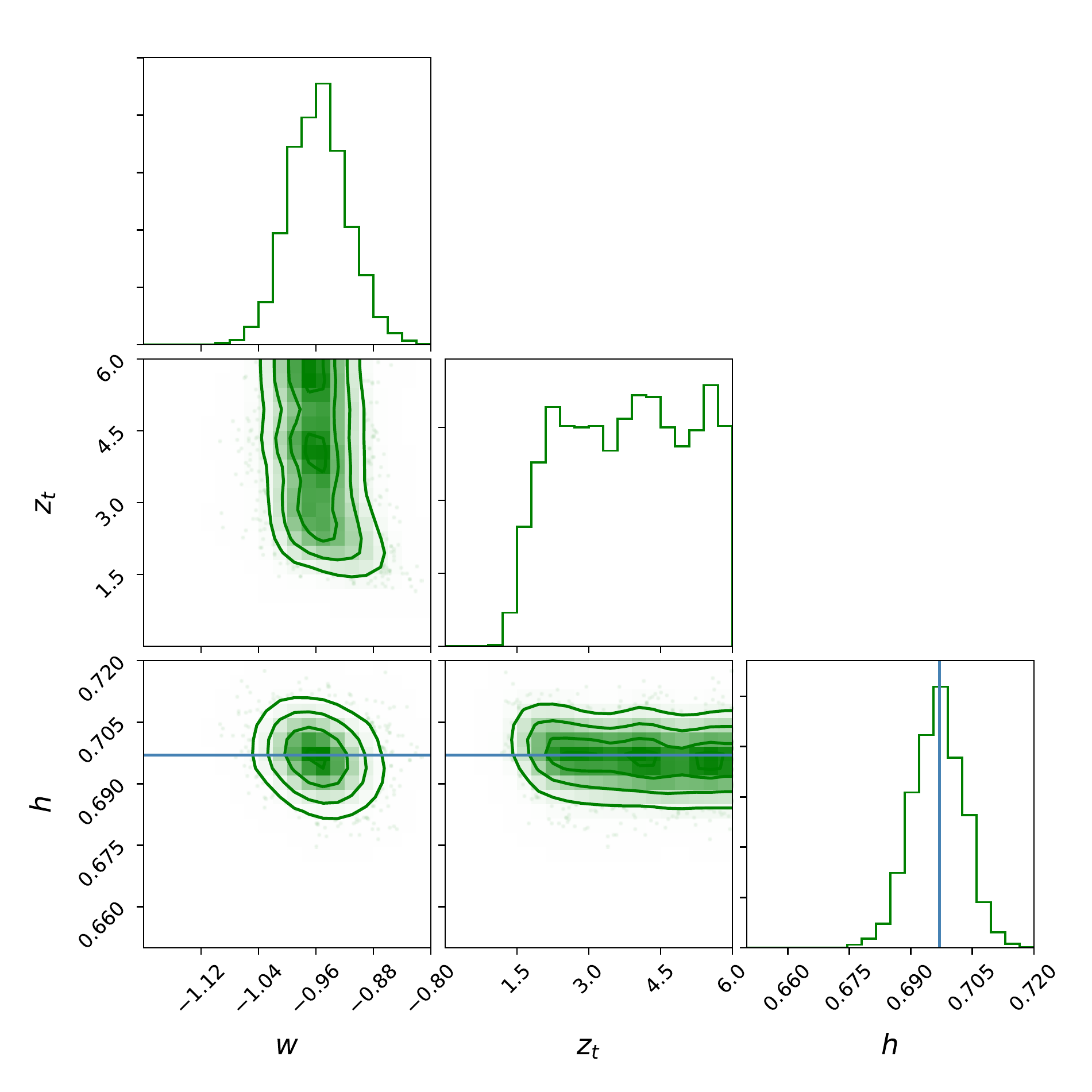}
\caption{\label{fig:triang_ztrans_withEos_h02} CMB/BAO/SN~Ia and local Hubble constant constraints on a phenomenological dark energy model for which at $z>z_t$, the dark energy density is zero. The inferred Hubble constant is $h=0.699 \pm 0.008$, close to the $\Lambda$CDM value for the same combination of data, indicated by the vertical blue line.}
\end{figure}

We imagine a model with a dark energy component having a constant EoS $w$, but for which at redshifts higher than a transition redshift, $z_t$, the dark energy density suddenly becomes zero. Although not motivated by any fundamental theory, this model is of interest since it represents the extreme end of minimizing the expansion rate at $z>z_t$ for non-negative energy densities. As is evident from figure~\ref{fig:triang_ztrans_withEos_h02}, the model closely reproduces the $\Lambda$CDM result, giving $h=0.699 \pm 0.008$. This shows that it is improbable that any dark energy model with non-negative effective energy density can push the Hubble constant as measured from CMB/BAO/SN~Ia data to the locally measured value of $h\approx 0.73$.

\subsubsection{Negative dark energy density}
\begin{figure}
\centering
\includegraphics[width=15cm]{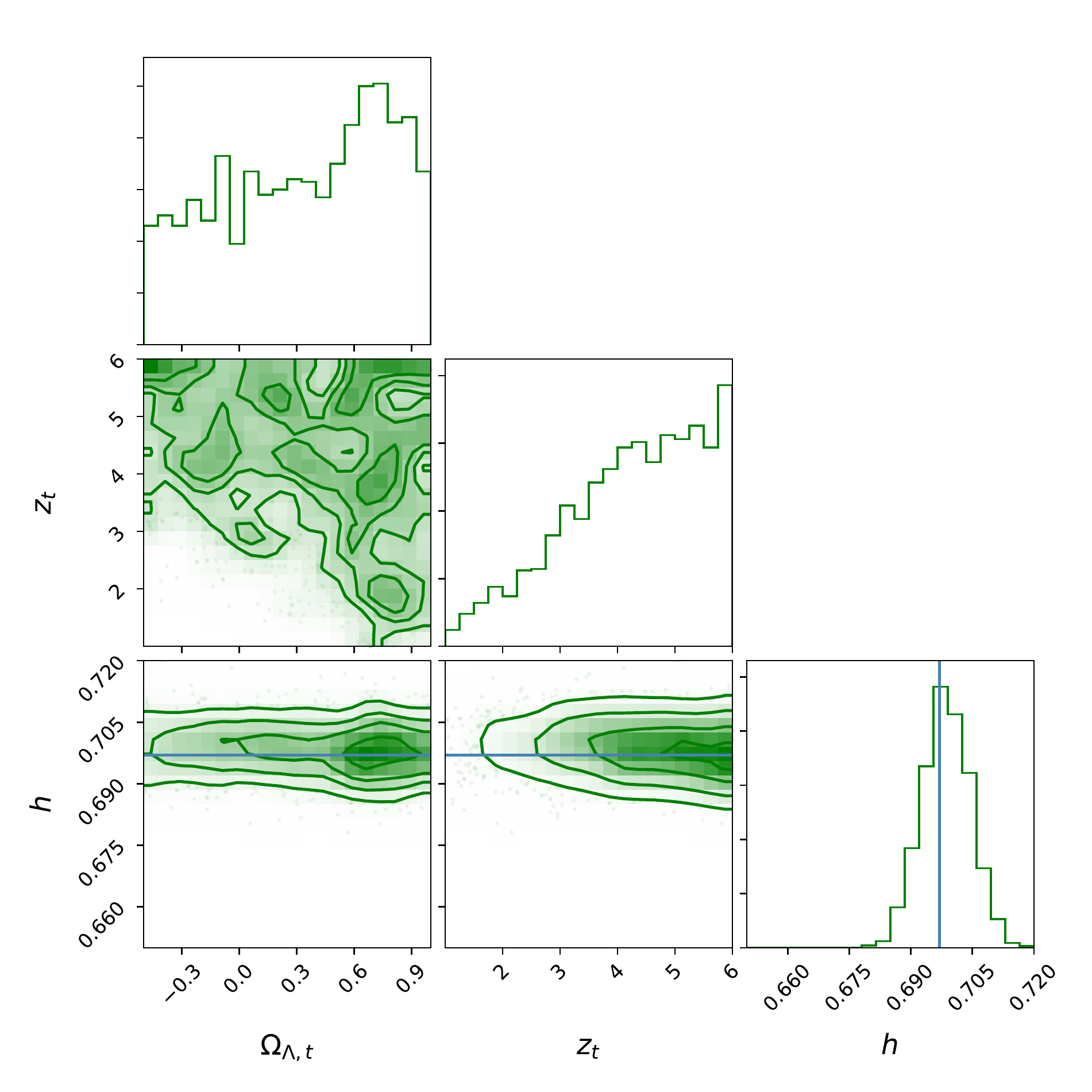}
\caption{\label{fig:triang_ztrans_withExoticDE_h0} CMB/BAO/SN~Ia and local Hubble constant constraints on a model for which at $z>z_t$, the cosmological constant can change value to $\Omega_{\Lambda,t}$. The data does not push for this possibility, and $h= 0.697\pm 0.007$. The vertical blue line is the $\Lambda$CDM value for the same combination of data.}
\end{figure}

Since dark energy models with positive energy densities fail to remove the Hubble tension, we next turn to a (again purely phenomenological) model, with a cosmological constant component which at a transition redshift, $z_t$, may change its value from $\ol$ to $\Omega_{\Lambda,t}$ (including negative densities). At least in principle, a negative constant at high redshifts could slow down the universal expansion enough to relieve the Hubble tension, without interfering with SN~Ia and BAO data at lower redshifts. Although such a model may at first seem very unphysical, we remind the reader that the quadratic bimetric model studied in section~\ref{sec:bimetric}, in fact allowed for the possibility of the effective dark energy density changing signature.
As is evident from figure~\ref{fig:triang_ztrans_withExoticDE_h0}, data does not support the presence of negative dark energy densities at high redshifts, and $h= 0.697\pm 0.007$.

\section{Early Dark Energy}
In this paper, dark radiation refers to any relativistic energy component with none or negligible electromagnetic interactions. Early dark radiation may affect CMB fluctuations at small scales by changing the damping scale \cite{Bernal2016}. In principle, this can be remedied by allowing for a scale dependence in the spectral tilt, $n$, but since the polarization spectrum is affected differently, polarization data can potentially break the degeneracy. Early dark radiation could also have an effect on the growth of perturbations. It can give an earlier freeze-out, and hence a greater abundance of helium and deuterium from BBN. In this work, we restrict our study to the effect of early dark radiation on the sound horizon size, $r_s$, at CMB decoupling, but we generalize the scope of dark radiation to include energy components with an arbitrary EoS, here denoted early dark energy (EDE). 

To increase the value of $h$ as inferred from CMB observations, we should decrease $r_s$, since for a smaller $r_s$ to subtend the same angle, we need a smaller $D_A(z_*)$ and thus a larger $h$. To decrease $r_s$, we can either decrease the sound speed $c_s$ or increase $H(z)$ at $z>z_*$. The sound speed is $c_s\approx 1/\sqrt{3}$ when photons dominate over baryons and approaches zero when baryons have started to dominate over photons, see equation~(\ref{eq:cs}). Since the baryon-to-photon ratio, and thus $c_s$, is tightly constrained by CMB fluctuations and BBN \cite{Ade:2015rim}, we focus on modifications to $H(z)$. 
In terms of the scale factor, the expansion rate is given by
\be
\left[\frac{H(a)}{100\,{\rm km/s/Mpc}}\right]^2=\frac{\wrad}{a^4}+\frac{\wm}{a^3}+\frac{\wk}{a^2}+\frac{\wde}{f(a)}.
\ee
where
\be
f(a)= \exp\left[3\int_a^1 \frac{[1+w_{\rm DE}(a)]da}{a}\right],
\ee
and 
\be
\wrad = \wg (1 + 0.2271 N_{\rm eff}).
\ee
$N_{\rm eff}$ is the effective number of neutrino species (with three neutrino species giving $N_{\rm eff}=3.046$). The density $\wrad$ is basically an umbrella for any kind of dark radiation with
\be
\omega=N_{\rm eff}\frac{7}{8}\left(\frac{4}{11}\right)^{4/3}\wg \approx 0.22711 N_{\rm eff}\wg.
\ee
Increasing $N_{\rm eff}$ makes the early Universe expand faster, $r_s$ smaller and $h$ larger since 
we need $D_A$ to be smaller. Note that if the additional contribution to $\wrad$ is to come from additional neutrino species, they have to be massless since otherwise, they will effectively add also to $\wm$ when becoming non-relativistic. The matter density is the sum of CDM, baryons and massive non-relativistic neutrinos 
\be
\wm=\omega_{\rm CDM}+\wb+\frac{\sum m_\nu}{94\,{\rm eV}}.
\ee
Increasing the neutrino mass is a well-motivated and non-controversial operation. However, the larger the neutrino mass, the lower the $h$ since neutrinos with masses below 1 eV will only become non-relativistic after decoupling and will thus affect the expansion velocity thereafter by adding to the matter density. This larger matter density can then be compensated for by a smaller Hubble constant, i.e., increasing the Hubble tension. 

\subsection{Dark radiation}
We first fit for an additional density component at pre-decoupling epochs with $\wede=1/3$, i.e. dark radiation $\Omega_{\rm DR}$. For a density of $\Omega_{\rm DR}=10^{-5}$ (one order of magnitude less than the ordinary radiation density), we find that the combination of local CMB, BAO and SN~Ia data gives $h =  0.737 \pm 0.00718$, in good agreement with the locally inferred value.

\begin{figure}
\centering
\includegraphics[width=11cm]{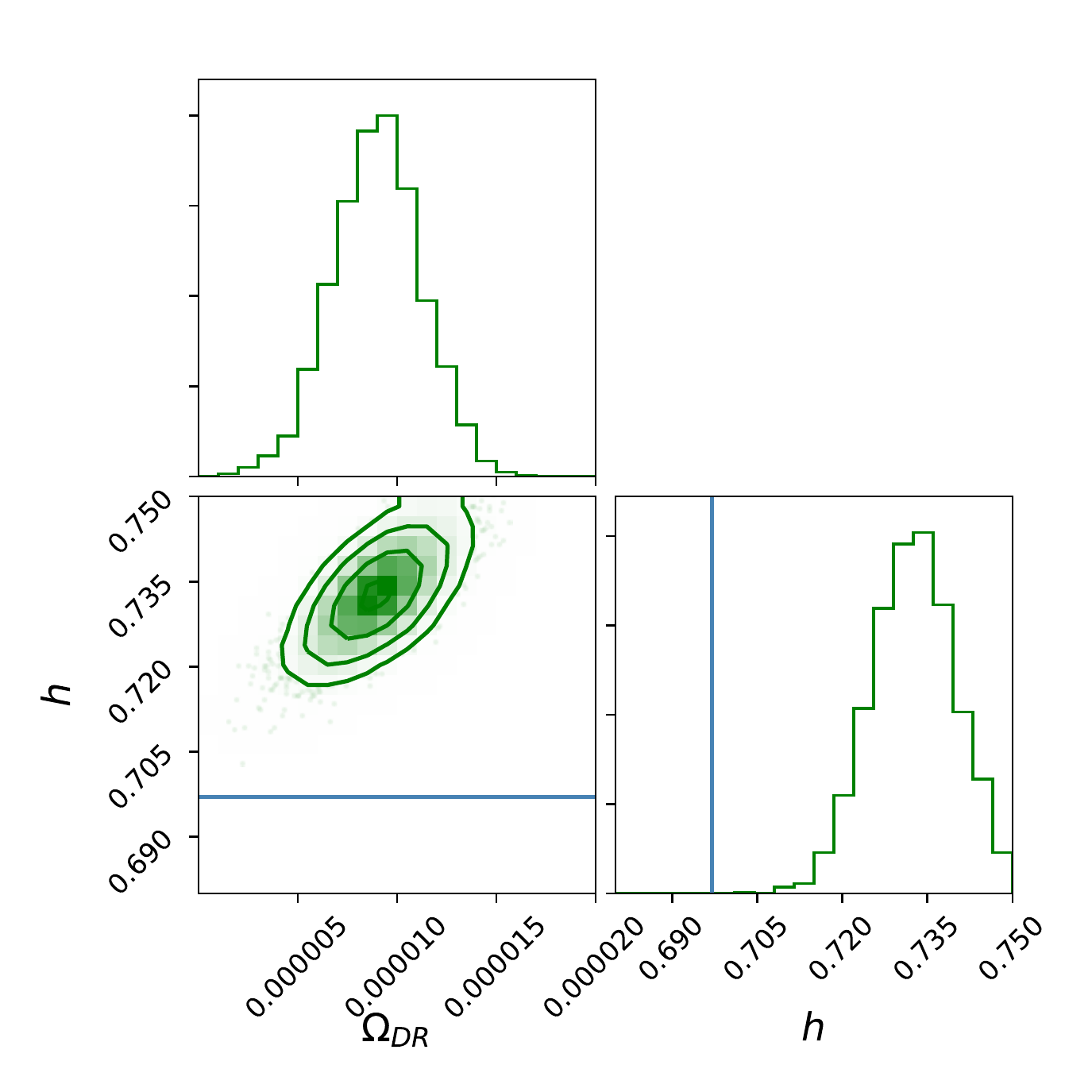}
\caption{Constraints on $h$ and $\Omega_{\rm DR}$ for a combination of CMB, BAO, SN~Ia and local $h$ data. The Hubble tension is solved by an additional $\sim 10\,\%$ relativistic energy density in the form of dark radiation. The vertical blue line is the $\Lambda$CDM value for the same combination of data.}
\label{fig:omega_dr}
\end{figure}

We next derive constraints on $\Omega_{\rm DR}$ using a uniform prior of [0., 0.01]. Since the shift parameter $l_A$ depends on the product of the Hubble constant and the sound horizon size, the derived value of $h$ is degenerate with $\Omega_{\rm DR}$. Hence, for further analyses, we continue combining CMB, BAO and SN~Ia data with the local prior on $h$. This constrains $\Omega_{\rm DR}$ and $h$ as shown in Figure~\ref{fig:omega_dr}. The best fit $\Omega_{\rm DR}$ is $8.79 \cdot 10^{-6} \pm 3.7 \cdot 10^{-6}$, consistent with 10$^{-5}$ and $h = 0.737 \pm 0.017$. The required additional density being $\sim 10\,\%$ of the total radiation density is marginally consistent with the bound from BBN discussed in section~\ref{sec:BBN}.

\subsection{The equation of state of early dark energy}
Additional early dark energy in the form of massless particles not interacting electromagnetically (i.e. dark radiation), will have an EoS of $w=1/3$. We next investigate to what extent observations can constrain a more general additional early dark energy EoS, $\wede$. If early dark energy is to increase the expansion rate at $z>z_*$, we expect negative $\wede$ to be ruled out since this would also increase the expansion rate at $z<z_*$, counteracting the effect at $z>z_*$. Also, if $\wede>1/3$, early dark energy will dominate the energy budget at very high redshifts and should be disfavoured by BBN constraints.   

\begin{figure}
\centering
\includegraphics[width=15cm]{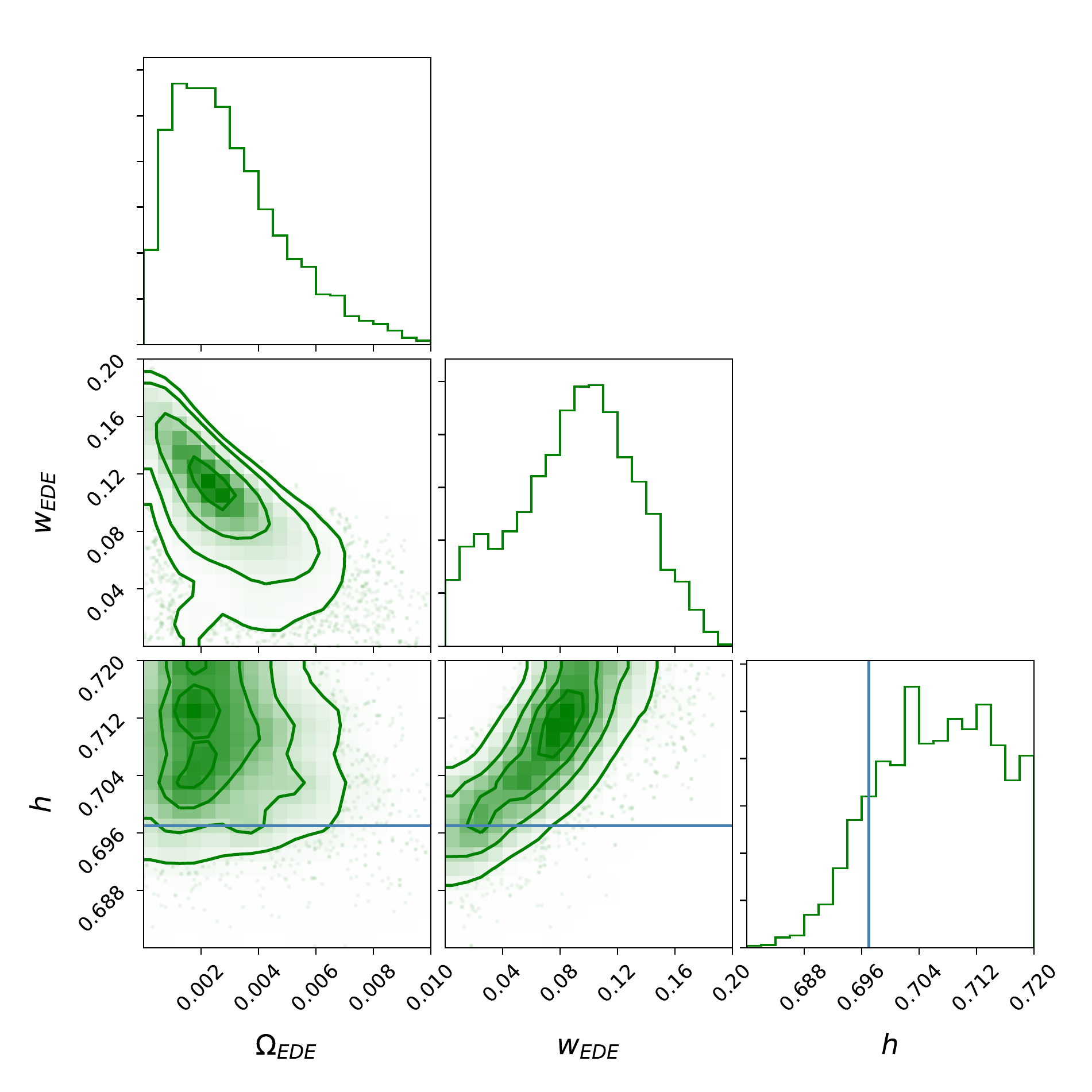}
\caption{\label{fig:triang_cmbbaosn_general_exoticDE_eosPossmall_h0} CMB/BAO/SN~Ia and local Hubble constant measurement constraints on a model with an additional early dark energy component with a constant EoS, $\wede$. The Hubble constant is $h=0.714\pm 0.0143$, in good agreement with the value inferred from local measurements only. The vertical blue line is the $\Lambda$CDM value of $h$ for the same combination of data.} 
\end{figure}

We assume a uniform prior on a constant $\wede$, [0, 0.4] and fit for the energy density $\Omega_{\rm EDE}$ and $\wede$. The resulting $\Omega_{\rm EDE}$, $\wede$ and $h$ are presented in Figure~\ref{fig:triang_cmbbaosn_general_exoticDE_eosPossmall_h0}. As predicted, $0<\wede<1/3$ is preferred. At $68\,\%$ CL, $\wede=0.086^{+0.04}_{-0.03}$, ruling out $\wede=1/3$ at more than $5\,\sigma$ CL.
The density for the additional EDE term change exponentially with the EoS, with more positive values of $\wede$ favouring smaller $\Omega_{\rm EDE}$. This is reflected in $\Omega_{\rm EDE}=0.0029 \pm 0.0019$ compared to the much smaller preferred value of $\Omega_{DR}$. 
The Hubble constant is $h=0.714 \pm 0.014$, within $1\,\sigma$ of the value inferred from local measurements only. 


\section{Bayesian evidence}\label{sec-bayesian}
In the previous sections, we have investigated to what degree additions and modifications to the $\Lambda$CDM model can push the Hubble constant as derived from CMB, BAO and SN~Ia data towards higher values more compatible with the locally inferred value. Our results indicate that early additional dark energy is more successful in doing this compared to late time dark energy, since BAO and SN~Ia data severely constrains any late time modifications to $\Lambda$CDM. In this section, we aim to quantify these results further. 

Still taking the locally measured value of $h$ as an unbiased measurement, we employ a Bayesian analysis in order to quantify the extent to which the additions and modifications studied are observationally preferred or not compared to the $\Lambda$CDM model. In short, we want to understand if the possible Hubble tension relief is worth the price of extra complexity introduced in the models. 

The Bayesian evidence, $Z$, is given by the integral of the prior times the likelihood over the entire parameter space of the model 
\begin{equation}
Z = \int L \pi d \theta .
\label{eq:evidence}
\end{equation}
Models for which the prior and the likelihood overlap to large extent will have large evidence. Taking the ratio of the Bayesian evidence for two models, we can quantify to what degree one model is preferred before the other. In practice, one often uses the logarithm of the evidence ratio for two models $Z_1$ and $Z_2$ and define
\begin{equation}
\Delta\ln Z\equiv \ln Z_1-\ln Z_2=\ln\frac{Z_1}{Z_2}.
\label{eq:logevidence}
\end{equation}
\begin{table*}
\caption{Priors on the free parameters of the models tested in this study. We use linear priors on all parameters.}
\begin{tabular}{|c|c|c|c|}
\hline
Model Parameter & Prior & Model\\
\hline
$\om$ & U[0, 1] & All\\
$B_1$ & U[0, 5] & Bimetric gravity\\	
$B_2$ & U[-100, 1.4] & Bimetric gravity\\	
$w$ & U[-2, 1] & (Dis)-appearing dark energy\\
$z_t$ & U[0, 6] & (Dis)-appearing dark energy and Negative dark energy \\
$\Omega_{\Lambda,t}$ & U[-2, 1] & Negative dark energy\\
$\Omega_{DR}$ & U[0, 0.00003] & Dark radiation \\
$\Omega_{\rm EDE}$ & U[0, 0.01]& Early dark energy \\
$w_{\rm EDE}$ & U[0, 0.4] & Early dark energy \\
$h$ & U[0.5, 1] & All\\
\hline
\end{tabular}
\label{tab:prior}
\end{table*}

The prior ranges for the model parameters employed in this paper are given in Table~\ref{tab:prior}. We use linear priors for all parameters. 

\begin{table}
\centering
\caption{The logarithm of the Bayesian evidence ($\ln Z$) for the models tested in this study and their comparison to the standard $\Lambda$CDM paradigm, $\Delta \ln Z\equiv \ln Z -\ln Z_{\Lambda{\rm CDM}}$. Some late time dark energy scenarios, like the disappearing dark energy model, are weakly disfavoured based on the $\Delta\ln Z$, whereas models with additional pre-CMB decoupling energy fare slightly better than the $\Lambda$CDM model.}
\begin{tabular}{|c|c|c|}
\hline
Model & ln $Z$ & $\Delta \mathrm{ln}\,Z$\\
\hline
$\Lambda$CDM & -48.1 & $\ldots$   \\
Linear bimetric gravity& -50.31 & 2.21 \\
Quadratic bimetric gravity& -48.1 & 0.0 \\
(Dis-)appearing dark energy & -51.4 & 3.3\\
Negative dark energy & -48.9 & 0.8 \\
Dark radiation & -46.8 & -1.3 \\
Early dark energy & -47.6 & -0.5 \\
\hline
$\Lambda$CDM (1$\%$ $H_0$) & -54.1 & $\ldots$   \\
Early dark energy (1$\%$ $H_0$) & -48.4 & -5.7   \\

\hline
\end{tabular}
\label{tab:evidence}
\end{table} 

In table~\ref{tab:evidence}, we report the $\ln Z$ and the difference relative to the fiducial $\Lambda$CDM model. The input data for all the models include the local prior on $h$. Despite their additional parameters, some of the late time dark energy descriptions are not disfavoured by data (e.g., the quadratic bimetric and the negative cosmological constant models), although they deviate little from their $\Lambda$CDM limiting case. Others, like the linear bimetric and the disappearing dark energy models, are moderately disfavoured. 

The model with a pre-decoupling early dark energy density has a positive evidence relative to the $\Lambda$CDM model, consistent with the model solving the Hubble tension. For the dark radiation case ($w = 1/3$), we also find a marginally positive evidence relative to $\Lambda$CDM. 
We note that the Bayesian evidence is higher for the dark radiation case since the improvement in the fit for the case with a free EoS of the early dark energy term is not significant enough to merit the additional free parameter. 
However, as always when interpreting Bayesian evidence, one should be aware that the range and functional shape of the parameter priors will affect the outcome of the analysis. For example, in the case of dark radiation, $\ln Z$ decreases by 1 if the prior range is expanded to 10$^{-4}$.


\subsection{Forecasts for future precision local measurements}
Future efforts to reduce the uncertainties on $h$ aim for a 1$\%$ precision \cite{2012arXiv1202.4459S}. Here, we compute the Bayesian evidence for the different cases tested above assuming an uncertainty of $1\%$ on the current value of $h$ from local distance ladder estimates, i.e., $h=0.7324 \pm 0.0073$. 

We use the 1$\%$ prior on $h$ to predict the value of $\Delta\ln Z$ for the early dark energy model with a constant EoS, $\wede$, relative to the standard $\Lambda$CDM case. Note that the EDE model has two additional free parameters relative to the standard case. The value for $\Delta\ln Z$ is 5.7 which would decisively favour the model with the additional early time density component.  

\section{Summary and Discussion}\label{sec-disc}
The recent discovery of an optical counterpart to a gravitational wave source \cite{2017ApJ...848L..16S} has led to the use of gravitational wave events as ``standard sirens'' to evaluate cosmological parameters, e.g. the Hubble constant \cite{ligoH0}. Although the method has large uncertainties from just the single object, it is expected to give precise future constraints on $h$, independent of the distance ladder \cite[e.g.,][]{Nissanke2013}. Independent constraints can also be obtained from HII regions in galaxies \cite{2017arXiv171005951F}, though they are currently less precise than SN~Ia distances, or from surface brightness fluctuation (SBF) distances \cite{2018arXiv180106080C},  though they haven't been extended to the Hubble flow. Future measurements of $h$  using  secondary distance indicators, e.g. the tip of the red giant branch stars (TRGB) \cite{Beaton2016} and Mira variables \cite{2018arXiv180102711H} are designed to provide independent, precise estimates (see \cite{2018arXiv180100598C} for a review). 
Such a plethora of different methods to measure the Hubble constant makes it an exciting time to understand possible causes of discrepancies between measurements. 

We have investigated how the local and CMB inferred Hubble constant values can be made compatible by 
modifying and adding extra energy components to the $\Lambda$CDM model. 
Models allowing for phantom energy pushes the dark energy EoS towards very negative $w<-1$ and high $h$ values as inferred from CMB data only. However, BAO and SN~Ia data at $z\lesssim 1$ are very effective in pushing $w\to -1$ and thus $h$ to low values. 

Additional dark energy at $z>z_*$ have the effect of speeding up the expansion rate before CMB decoupling, giving a smaller $r_s$ and a higher inferred value of $h$. 
Including the local Hubble prior, the constant equation of state of an additional early dark energy density component is constrained to $\wede=0.086^{+0.04}_{-0.03}$ and the Hubble constant to $h=0.714\pm 1.4$. This result disfavours dark radiation with $w = 1/3$ at $> 5\,\sigma$ as the cause of relief of the Hubble tension, though the radiation case returns a higher $h$ value. However, due to the smaller prior parameter volume of the dark radiation model, compared to the the case of an arbitrary $\wede$, it is nevertheless weakly preferred in terms of Bayesian evidence.  

Modifications at $z<z_*$ are thus too constrained by BAO and SN~Ia data to effectively relieve the tension, whereas modifications at $z>z_*$ are moderately successful. In a Bayesian sense, current data are insufficient to rule in or out any of the models tested in this paper. 
However, given that the Hubble tension persists, future estimates of $h$ at the 1$\%$ level will be able to decisively determine if early dark energy models are favoured compared to $\Lambda$CDM. 




{\it Acknowledgements:} 
We would like to thank Rahul Biswas for stimulating discussions. 
\bibliographystyle{unsrt}
\bibliography{biblio}

\end{document}